# Commensurate-incommensurate phase transition in bilayer graphene

*Andrey M. Popov[1,*], Irina V. Lebedeva[2,3,4,†], Andrey A. Knizhnik[2,3,‡], Yurii E. Lozovik[1,4,§], and Boris V. Potapkin[2,3]*

[1]Institute of Spectroscopy, Fizicheskaya Street 5, Troitsk, Moscow Region, 142190, Russia,

[2]National Research Centre "Kurchatov Institute", Kurchatov Square 1, Moscow, 123182, Russia,

[3]Kintech Lab Ltd, Kurchatov Square 1, Moscow, 123182, Russia,

[4]Moscow Institute of Physics and Technology, Institutskii pereulok 9, Dolgoprudny, Moscow Region, 141701, Russia

**ABSTRACT**

A commensurate-incommensurate phase transition in bilayer graphene is investigated in the framework of the Frenkel-Kontorova model extended to the case of two interacting chains of particles. Analytic expressions are derived to estimate the critical unit elongation of one of the graphene layers at which the transition to the incommensurate phase takes place, the length and formation energy of incommensurability defects (IDs) and the threshold force required to start relative motion of the layers on the basis of dispersion-corrected density functional theory calculations of the interlayer interaction energy as a function of the relative position of the layers. These estimates are confirmed by atomistic calculations using the DFT-D based classical potential. The possibility to measure the barriers for relative motion of graphene layers by the study of formation of IDs in bilayer graphene is discussed.

[*] am-popov@isan.troitsk.ru

[†] lebedeva@kintechlab.com

[‡] knizhnik@kintechlab.com

[§] lozovik@isan.troitsk.ru

## I. INTRODUCTION

In addition to zero-dimensional and one-dimensional carbon nanostructures, fullerenes and carbon nanotubes, a novel two-dimensional carbon nanostructure, graphene, was discovered recently[1]. Outstanding electrical and mechanical properties of graphene have been utilized in flexible transparent electrodes[2]. Stiff and flexible graphene oxide paper[3] is promising for the use in fuel cell and structural composite applications. Moreover, a number of graphene-based nanoelectromechanical systems were proposed recently. A nanoresonator based on flexural vibrations of suspended graphene was implemented[4]. The experimentally observed self-retracting motion of graphite, i.e. retraction of graphite flakes back into graphite stacks on their extension arising from the van der Waals interaction, led to the idea of a gigahertz oscillator based on the telescopic oscillation of graphene layers[5]. Nanorelays based on the telescopic motion of nanotube walls[6,7] were realized experimentally. Furthermore, a mass nanosensor based on the small translational vibrations of nanotube walls was considered[8,9]. By analogy with these devices, a nanorelay based on the telescopic motion of graphene layers and a mass nanosensor based on the small translational vibrations of graphene layers can be proposed[10]. For all these applications and for understanding of fundamental properties of graphene, investigation of mechanical properties of few-layer graphene associated with the relative displacement of the layers is of high importance. The interaction between walls of carbon nanotubes and between the nanotubes themselves is similar to the interaction between graphene layers. Therefore, investigation of the interlayer interaction and relative displacement of graphene layers is also of interest for the development of nanotube-based nanoelectromechanical systems[6–9,11] and for studies of the interaction between carbon nanotubes in fibers and yarns[12].

The mechanical properties of few-layer graphene associated with the relative displacement of the layers have been so far poorly studied experimentally. The value of the shear strength for graphite determined from the only known experiment[13] was claimed to be related to macroscopic structural defects of the graphite sample. According to the results of calculations[14–17], the shear strength for relative motion of carbon nanotubes walls depends on chiral indices of the walls and differs by orders of magnitude for different pairs of the walls. Therefore, interpretation of the experimental data on the shear strength for carbon nanotubes[18–20], which show significant scatter, is not possible without the information on the chiral indices of the nanotube walls. The values of the shear strength for graphene and carbon nanotubes are determined by the characteristics of the potential relief of the interaction energy of graphene layers and nanotube walls. The potential relief of the interlayer interaction energy was investigated in experiments[21,22] for a graphene flake moved on a graphite surface by the tip of the friction force microscope. However, in these experiments, only a small region of the potential energy relief was accessible. Based on our study, we propose the way to experimentally determine the barrier for relative motion of graphene layers by observation of a structural transition in bilayer graphene upon stretching or compression of one of the layers. The knowledge of this barrier is particularly valuable for the verification of *ab initio* studies of relative motion of graphene layers and carbon nanotube walls and for the development of a classical potential for simulation of related phenomena.

The weakness of van der Waals forces between graphene layers makes possible the relative displacement of graphene layers in graphite[5]. Small graphene flakes can be easily moved as a whole across a graphite surface by the microscope tip[21,22]. However, the relative displacement of a single large-scale graphene layer on another graphene layer or a graphite surface should be accompanied by the deformation of the layers. As a result, the movable layer can become incommensurate with the underlying layer. Here we for the first time consider a commensurate-incommensurate phase transition[23–27] in bilayer graphene with one tension layer. At small unit elongations of the tension layer, the layers are commensurate, i.e. they have equal elementary unit cells, and the system is in the commensurate phase. However, the elastic energy of the system increases with stretching or compression of the tension layer. At some critical unit elongation, the necessity for lowering the elastic energy of the system results in the relative displacement of the free layer and formation of the first incommensurability defect (ID), that is the transition to the incommensurate phase takes place. The number of IDs increases with the further increase of the unit elongation and the incommensurate phase has a structure of alternating long nearly commensurate regions and short IDs.

To investigate the commensurate-incommensurate phase transition in bilayer graphene in the present paper we modify the Frenkel-Kontorova (FK) model[23]. Such transitions have been generally considered in systems consisting of a harmonic chain of particles in a periodic potential[23–27], i.e. the substrate was assumed rigid. In papers[28,29], it was shown that the interaction of free walls of double-walled carbon nanotubes of similar rigidity can be effectively described using the same FK model. To describe this phase transition in bilayer graphene upon stretching or compression of one of the layers, we extend the one-dimensional FK model to the case of two nonlinearly interacting harmonic chains one of which is under the strain and another one is free. We use the modified FK model to obtain analytic expressions for the length and formation energy of IDs and for the critical unit elongation of bilayer graphene. We also estimate the threshold force required to start sliding of one graphene layer on another. These analytic estimates are verified on the basis of atomistic calculations. The developed theory is also applicable to other interacting one-dimensional nanoobjects, for example, to nanotube walls, one-dimensional nanostructures inside nanotubes[30–32], and interacting parallel nanotubes[11].

We show that the critical unit elongation and the length of IDs in bilayer graphene depend on the barrier for relative motion of graphene layers. Thus, we propose that experimental measurements of the critical unit elongation and the length of IDs would provide the correct value of the barrier for relative motion of graphene layers. Similar to rotational stacking faults in few-layer graphene giving rise to Moiré patterns[33–35], IDs formed in bilayer graphene can be observed using HRTEM and STM. To estimate the critical unit elongation for bilayer graphene in the present paper we calculate the barrier for relative motion of graphene layers using the recent dispersion-corrected density functional theory (DFT-D) approach[36,37].

The paper is organized as follows. The results of the DFT-D calculations of the barrier for relative motion of graphene layers are discussed in Sec. II. In Sec. III, formation of the first ID in bilayer graphene is investigated using the modified FK model. The results of the atomistic calculations of the critical unit elongation for bilayer graphene are given in Sec. IV. Our conclusions are summarized in Sec. V.

## II. POTENTIAL RELIEF OF INTERLAYER INTERACTION ENERGY

To estimate the critical unit elongation for bilayer graphene we have performed the calculations of the potential relief of the interlayer interaction energy using the VASP code[38] with the generalized gradient approximation (GGA) density functional of Perdew, Burke, and Ernzerhof (Ref. 39) corrected with the dispersion term (PBE-D)[37]. The periodic boundary conditions are applied to a 4.271 Å x 2.466 Å x 20 Å model cell. The basis set consists of plane waves with the maximum kinetic energy of 800 eV. The interaction of valence electrons with atomic cores is described using the projector augmented-wave method (PAW)[40]. Integration over the Brillouin zone is performed using the Monkhorst-Pack method[41] with 24x36x1 k-point sampling. In the calculations of the potential energy relief, one of the graphene layers is rigidly shifted parallel to the other. Account of structure deformation induced by the interlayer interaction was shown to be inessential for the shape of the potential relief for the interaction between graphene-like layers, such as the interwall interaction of carbon nanotubes[16] and the intershell interaction of carbon nanoparticles[42,43].

From the potential relief of the interlayer interaction energy for bilayer graphene obtained on the basis of the DFT-D calculations (see FIG. 1a), it is seen that the minimum energy path for transition of the graphene layers between adjacent energy minima corresponds to the relative displacement of the layers in armchair directions. Therefore, formation of IDs in bilayer graphene should proceed by the relative displacement of the graphene layers in the armchair directions. The interlayer interaction energy along the path from one energy minimum to an adjacent one can be approximated by a cosine function (see FIG. 1b)

$$V(u) = 0.5W(1 - cos(2\pi u)),\quad(1)$$

where $u$ is the dimensionless relative displacement of the graphene layers in the armchair direction measured relative to the bond length of graphene $l_0 = 1.42$ Å. The parameter $W$ per atom of one of the layers is fitted to be $W = 2.10$ meV/atom. The relative root-mean-square deviation of this approximation from the minimum energy path obtained from the DFT-D calculations is found to be $\delta U / W = 0.028$.

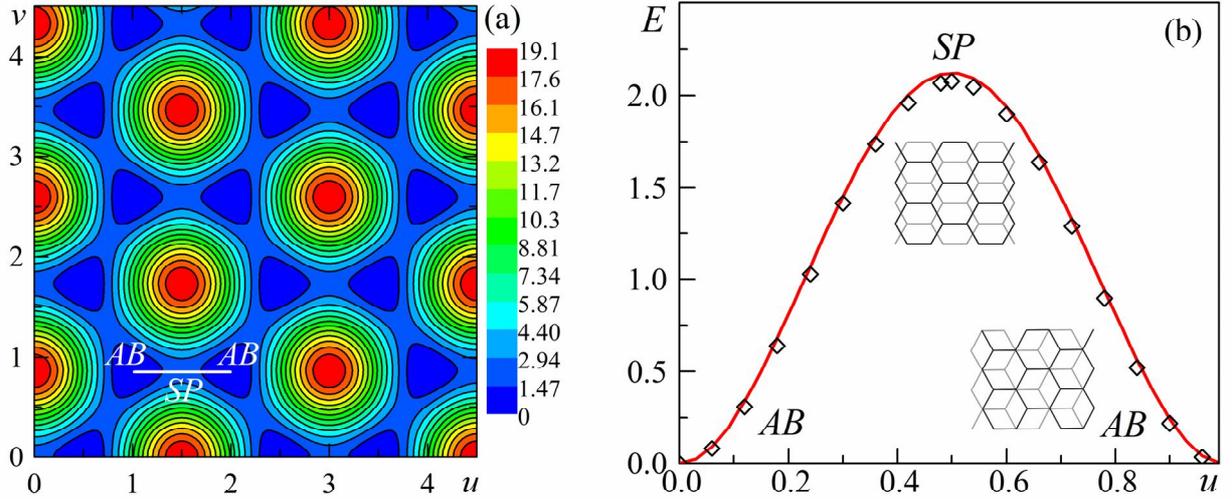

FIG. 1. (a) Calculated interlayer interaction energy of bilayer graphene (in meV/atom) as a function of the relative position of the graphene layers $u$ and $v$ (measured relative to the bond length of graphene $l_0 = 1.42$ Å; $u$ and $v$ axes correspond to the armchair and zigzag directions, respectively). The minimum energy path corresponding to transition between adjacent energy

minima is shown by the white line. (b) Calculated interlayer interaction energy of bilayer graphene (in meV/atom) at the equilibrium interlayer spacing as a function of the relative displacement $u$ of the layers along the minimum energy path. The solid line shows the approximation of the data obtained using the DFT-D calculations with the cosine function (1). The energy is given relative to the global energy minimum.

## III. TWO-CHAIN FRENKEL-KONTOROVA MODEL

To investigate the possibility of the commensurate-incommensurate phase transition in bilayer graphene with one of the layers being stretched or compressed along the armchair direction, we consider a system consisting of two harmonic chains comprising the same number of particles $N \gg 1$ (see FIG. 2). It is assumed that each "particle" in the model corresponds to a strip of graphene directed perpendicular to the elongation (along the zigzag direction) and of the width equal to the bond length. Let the length of chain II be fixed at some value $L = Nl$, which is only slightly different from the equilibrium length $L_0 = Nl_0$ of chain I in the case when it is isolated and not interacting with chain II ($|(L-L_0)/L| \ll 1$). Chain I is free.

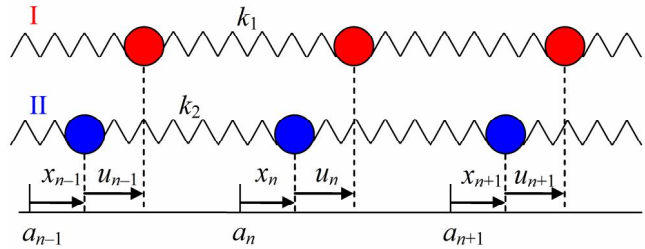

FIG. 2. Frenkel-Kontorova model for two interacting chains of particles.

The total energy of the system is given by the sum of the elastic energy $U_{el}$ of the chains and the energy of their interaction $U_{int}$, $U = U_{el} + U_{int}$. The elastic energy of the chains can be found as

$$U_{el} = \sum_{n=-N/2}^{N/2} \left( \frac{k_1 l^2 (\delta_1 + x_n + u_n - x_{n-1} - u_{n-1})^2}{2} + \frac{k_2 l^2 (\delta_2 + x_n - x_{n-1})^2}{2} \right), \qquad (2)$$

where $k_1$ and $k_2$ are the elastic constants of the springs of chains I and II, respectively (for graphene layers $k_1 = k_2 = k$; for a rigid substrate $k_2 \to \infty$), $x_n$ correspond to the displacements of the particles of chain II relative to their equidistant positions $a_n = nl$ in the isolated chain, $u_n$ correspond to the displacements of the particles of chain I relative to the particles of chain II, $\delta_1$ and $\delta_2$ are the relative elongations of the springs of chains I and II, respectively, if their lengths equal $l$ (we consider here the general case of different equilibrium lengths of isolated chains I and II). The quantities $x_n$, $u_n$, $\delta_1$ and $\delta_2$ are dimensionless and measured relative to $l$. The interaction of the chains is described by the sum of the cosine functions $U_{int} = \sum_n V(u_n)$ (see Eq. (1)).

Within the continuum approximation, summation over the particles becomes an integration over $n$. Taking into account that $\int_{-N/2}^{N/2} (dx/dn)\,dn = 0$ as the length of chain II is fixed, we get the total energy of the system in the form $U = U_{com} + \Delta U$, where $U_{com} = N(k_1 l^2 \delta_1^2 + k_2 l^2 \delta_2^2)/2$ corresponds to the energy of the commensurate state and $\Delta U = U_1 + U_2$, where

$$U_1 = \frac{1}{2} \int_{-N/2}^{N/2} \left( k_1 l^2 \left( \frac{dx}{dn} + \frac{du}{dn} \right)^2 + k_2 l^2 \left( \frac{dx}{dn} \right)^2 + 2V(u) \right) dn, \qquad (3)$$

$$U_2 = \int_{N/2}^{N/2} k_1 l^2 \delta_1 \frac{du}{dn} dn = k_1 l^2 \delta_1 \left( u\left(\frac{N}{2}\right) - u\left(-\frac{N}{2}\right) \right) = k_1 l^2 \delta_1 \Delta u, \tag{4}$$

determines the relative energy of the incommensurate state.

Introducing a variable $\varphi = x + k_1 u / (k_1 + k_2)$, the energy $U_1$ can be rewritten as

$$U_1 = \int_{-N/2}^{N/2} \left( \frac{1}{2} K_1 l^2 \left(\frac{d\varphi}{dn}\right)^2 + \frac{1}{2} K_2 l^2 \left(\frac{du}{dn}\right)^2 + V(u) \right) dn, \tag{5}$$

where $K_1 = k_1 + k_2$ and $K_2 = k_1 k_2 / (k_1 + k_2)$. Note that expression (4) is the same as in the case of the original FK model for a chain of particles on a rigid substrate[27]. In expression (5), the effective elastic constant $K_2$ is used instead of the elastic constant $k_1$ of the springs of chain I (Ref. 27). Furthermore, the first term in expression (5) is new. Both of these differences are related to deformation of the substrate (chain II), which is not taken into account in the original FK model.

Integration of the Euler-Lagrange equations $\delta U / \delta \varphi = 0$ and $\delta U / \delta u = 0$ satisfied at extrema of $\Delta U$ gives

$$\frac{1}{2} K_1 l^2 \left(\frac{d\varphi}{dn}\right)^2 = \varepsilon_1, \quad \frac{1}{2} K_2 l^2 \left(\frac{du}{dn}\right)^2 = V(u) + \varepsilon_2 \tag{6}$$

where $\varepsilon_1, \varepsilon_2 \geq 0$ are integration constants with units of energy. The case of $\varepsilon_2 > 0$ is considered in Appendix.

Formation of the first ID corresponds to the solutions with $\varepsilon_2 = 0$ and $\Delta u = \pm 1$. From the second of Eqs. (6), we obtain for the boundary condition $\Delta u = -1$ ($N \to \infty$)

$$u = -\frac{2}{\pi} \arctg \left( \exp \left( \pi n \sqrt{\frac{2W}{K_2 l^2}} \right) \right). \tag{7}$$

It is seen that the effective length of the ID is given by $l_{ID} = l \left( du/dn \big|_{n=0} \right)^{-1} = l^2 \sqrt{K_2/(2W)}$.

To satisfy the boundary condition $x(N/2) - x(-N/2) = 0$ we set the parameter $\varepsilon_1$ to be $\varepsilon_1 = k_1^2 l^2 / (2K_1 N^2)$. So the solution for $x$ finally takes the form

$$x = \frac{K_2}{k_2} \left( \frac{2}{\pi} arctg \left( exp \left( \pi \sqrt{\frac{2W}{K_2 l^2}} n \right) \right) - \frac{n}{N} - \frac{1}{2} \right). \tag{8}$$

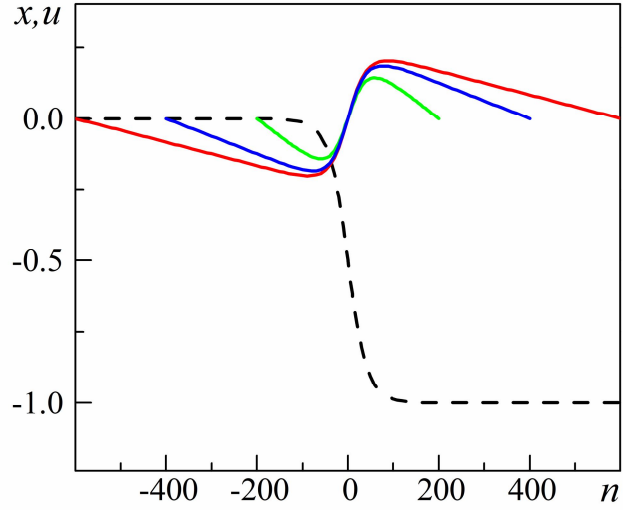

FIG. 3. Calculated structure of the first incommensurability defect: displacements $u(n)$ (dashed line) of the particles corresponding to the free graphene layer relative to the particles of the tension layer and displacements $x(n)$ (solid lines) of the particles corresponding to the tension layer relative to the perfect graphene lattice for bilayer graphene of length $L_0 \approx 57$, 114 and 170 nm ($N = 400$, 800 and 1200).

For a rigid substrate ($k_1 \ll k_2$), we have $K_2 \approx k_1$ and $x \ll 1$. Thus, the solution obtained is consistent with the solution of the original FK model. For bilayer graphene ($k_1 = k_2 = k$), $K_2 \approx k/2$ and $x \sim 0.2 \sim u$ at $|n| \sim l_{ID}/l$. Therefore, in this case, the displacements of atoms of the tension graphene layer from the perfect graphene lattice are comparable to the displacements of atoms of the free graphene layer (see FIG. 3). Note also that solution (7), (8) is derived without an assumption of equality of the equilibrium lengths of isolated chains I and II. This makes it applicable also to the case of systems consisting of interacting one-dimensional nanoobjects with slightly different lattice constants, such as few-walled nanotubes, crystals[30–32] and polymers inside nanotubes, and interacting parallel nanotubes[11].

The relative energy of the solutions corresponding to the boundary condition $\Delta u = \pm 1$ (see Eqs. (7) and (8)) can be found from Eqs (5), (6)

$$\Delta U = \int_0^1 \sqrt{2K_2 l^2 V(u)} du \pm k_1 l^2 \delta_1 + k_1^2 l^2 / (2K_1 N). \tag{9}$$

Formation of the first ID in the commensurate phase occurs when $\Delta U = 0$. Therefore, the critical unit elongation is given by

$$\delta_c = \mp \frac{1}{k_1 l^2} \int_0^1 \sqrt{2K_2 l^2 V} du \mp \frac{k_1}{2K_1 N} = \mp \frac{2\sqrt{2K_2 l^2 W}}{\pi k_1 l^2} \mp \frac{k_1}{2K_1 N}. \tag{10}$$

At $N \gg k_1 / (2K_1 \delta_{c0})$ the critical unit elongation reaches $\delta_{c0} = U_{ID} / (k_1 l^2)$, where $U_{ID} = 2\sqrt{2K_2 l^2 W}/\pi$ is the energy of an isolated ID at the unit elongation $\delta_1$ equal to zero.

To estimate the threshold force required to start sliding of one graphene layer on another along the armchair direction we consider the FK model for two interacting chains of particles in the case when the chains are identical ($k_1 = k_2 = k$) and free ($\delta_1 = \delta_2 = 0$) and a stretching force $F$ is applied to the last particle of chain I and to the first particle of chain II. The energy of such a system is given by

$$U = U_1 - Flu(N/2) - Fl(x(N/2) - x(-N/2)), \tag{11}$$

where $U_1$ is given by expression (3). It is seen that the solution in this case should also satisfy the Euler-Lagrange equations (6). At equilibrium, the balance of the forces for chain I with account of the second of the Euler-Lagrange equations (6) yields

$$Fl = \int_{-N/2}^{N/2} \frac{\partial V}{\partial u} dn = \int_{-N/2}^{N/2} K_2 l^2 \frac{d^2 u}{dn^2} dn = K_2 l^2 \left( \frac{du}{dn}\left(\frac{N}{2}\right) - \frac{du}{dn}\left(-\frac{N}{2}\right) \right). \tag{12}$$

From Eq. (7), it follows that the equilibrium is possible for $F < F_0$, where the threshold force $F_0$ required to start relative motion of graphene layers is seen to be $F_0 \approx \pi W N / l$ for $N \ll l_{\text{ID}}/l$ and $F_0 \approx K_2 l^2 / l_{\text{ID}} = \sqrt{2WK_2}$ for $N \gg l_{\text{ID}}/l$. So for large layers the threshold force does not depend on the overlap length of the layers in the direction of the force.

Let us use the derived expressions to estimate the critical unit elongation for bilayer graphene with one layer being stretched or compressed along the armchair direction, the energy and length of IDs in graphene and the threshold force required to start relative motion of the layers. We use approximation (1) for the interlayer interaction energy. The Young modulus for graphene was measured[44] to be $Y = 1.0 \pm 0.1$ TPa for the effective thickness of graphene $t = 3.35$ Å. The effective elastic constant per unit width can, therefore, be estimated as $k = Yt/l \approx 15$ eV/Å$^3$. The barrier for relative motion of graphene layers is $W = 1.13$ meV/Å per one particle of the considered model with a unit width along the zigzag direction perpendicular to the elongation. From the equations derived above, we find for bilayer graphene $l_{\text{ID}} = 0.5 l^2 \sqrt{k/W} \approx 12$ nm, $U_{\text{ID}} = 2\sqrt{kl^2 W}/\pi = 0.12$ eV/Å (per unit width of the layers perpendicular to the elongation) and $\delta_{c0} = U_{\text{ID}}/(kl^2) = 3.9 \cdot 10^{-3}$, respectively. The threshold force required to start relative motion of graphene layers in the armchair direction is estimated to be $F_0 \approx \sqrt{kW} \approx 0.13$ eV/Å² $\approx 0.21$ nN/Å (per unit width). As the critical unit elongation $\delta_{c0}$ and the length of IDs $l_{ID}$ depend on the barrier for relative motion of graphene layers $W$, experimental measurements of $\delta_{c0}$ and $l_{ID}$ would allow to determine this barrier. Similar to rotational stacking faults in few-layer graphene giving rise to Moiré patterns[33–35], IDs formed in bilayer graphene can be observed using HRTEM and STM.

## IV. ATOMISTIC CALCULATIONS

To confirm the obtained analytic estimates we have performed atomistic calculations of the energies of bilayer graphene in the commensurate state and with a single ID as functions of the unit elongation of one of the layers. An in-house MD-kMC (Molecular dynamics – kinetic Monte Carlo) code[45] is used. The covalent carbon-carbon interactions in the layers are described by the Brenner potential[46]. The van der Waals interaction of the layers is described using the potential developed recently on the basis of the DFT-D calculations[10]. The cutoff distance of this van der Waals potential is taken equal to 12 Å.

The system consisting of two graphene layers at the equilibrium interlayer spacing of 3.374 Å is considered. At equilibrium, both of the layers have length of $L = 426$ Å along the armchair direction ($N = 300$ in the one-dimensional FK model). The periodic boundary conditions are applied along the perpendicular zigzag direction. If the layers are not stretched, the size of the model cell along this direction is 29.5 Å. As one of the layers is stretched, the size of the model cell along the zigzag direction is decreased according to the Poisson ratio. Two structures of bilayer graphene corresponding to the commensurate state and to the incommensurate state with a single ID are considered. The energy optimization is performed for both of the structures using the conjugated gradient method.

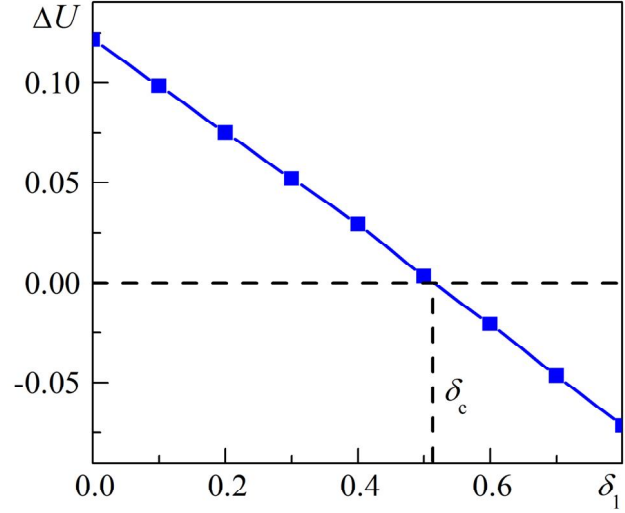

FIG. 4. Difference in the energies of bilayer graphene in the incommensurate state with a single ID and in the commensurate state $\Delta U$ (in meV/Å, per unit width of the layers perpendicular to the elongation) as a function of the unit elongation of the tension layer $\delta_l$ (in %) obtained from the atomistic calculations.

The atomistic calculations show that the relative energy of the structure with a single ID has a linear dependence on the unit elongation of the tension layer and goes to zero at the critical unit elongation of $\delta_c = 5.13 \cdot 10^{-3}$ (see FIG. 4). For the Brenner potential, we find the equilibrium bond length in bilayer graphene to be 1.420 Å. The effective elastic constant per unit width is calculated to be $k \approx 12 \, \text{eV/Å}^3$. The Poisson ratio is found to be 0.19. These values are in agreement with the experimental data[44] and the results of *ab initio* calculations[47]. Based on formula (10), we estimate the critical unit elongation for the considered system to be $\delta_c = 5.23 \cdot 10^{-3}$. It is seen that the difference in the critical unit elongations obtained using the atomistic calculations and Eq. (10) is within 2%.

The positions of atoms in the structure with a single ID only slightly deviate from solution (7), (8) upon relaxation. The root-mean-square deviations of $x$ and $u$ calculated for the system with a single ID from solution (7), (8) are only $1.5 \cdot 10^{-3}$ and $1.0 \cdot 10^{-2}$, respectively.

## V. CONCLUSION

We extend the Frenkel-Kontorova model to the case of two interacting harmonic chains of particles. Using this modified model, we for the first time investigate the commensurate-incommensurate transition in bilayer graphene. On the basis of the DFT-D calculations of the potential relief of the interlayer interaction energy for bilayer graphene, we predict the critical unit elongation of one of the layers of bilayer graphene along the armchair direction at which the formation of the first ID occurs to be about 0.39%. The length of IDs is estimated to be 12 nm. The energy of an isolated ID in the system of the infinite length is found to be 0.12 eV/Å (per unit width of the layers perpendicular to the elongation) at the zero unit elongation. The threshold force required to start relative motion of graphene layers in the armchair direction is shown to be independent of the overlap length of the layers in the direction of the force for large layers and is estimated to be 0.21 nN/Å (per unit width). The estimates are confirmed by the atomistic calculations using the DFT-D based classical potential. We propose that experimental measurements of the critical unit elongation for bilayer graphene and of the length of IDs can provide the correct value of the barrier for relative motion of graphene layers. The theory developed in the present paper is also applicable to other interacting one-dimensional nanoobjects, for example, to nanotube walls, one-dimensional nanostructures inside nanotubes[30–32] and interacting parallel nanotubes[11].

**ACKNOWLEDGEMENTS**

This work has been partially supported by the RFBR grants 11-02-00604 and 10-02-90021-Bel. The atomistic calculations are performed on the SKIF MSU Chebyshev supercomputer and on the MVS-100K supercomputer at the Joint Supercomputer Center of the Russian Academy of Sciences.

**APPENDIX**

Let us consider the case when the parameter $\varepsilon_2 > 0$. From the second of the Euler-Lagrange equations (6), it follows that

$$dn = \pm \sqrt{\frac{K_2 l^2}{2(V(u) + \varepsilon_2)}} du. \tag{A1}$$

As the function $V(u)$ is periodic, the solution is also periodic for $\varepsilon_2 > 0$ and the distance between IDs is given by

$$\Gamma = \int_0^1 \sqrt{\frac{K_2 l^2}{2(V(u) + \varepsilon_2)}} du. \tag{A2}$$

At $\varepsilon_2 \to 0$ the distance between IDs $\Gamma \to \infty$, so that only one ID can be present in the system.

The value of the parameter $\varepsilon_1$ is found to be $\varepsilon_1 = k_1^2 l^2 \Delta u^2 / (2 K_1 N^2)$ from the first of the Euler-Lagrange equations (6) and the boundary condition $\Delta x = 0$.

For $\Gamma < N$, $\Delta u = \pm N / \Gamma$ and the expression for $\Delta U$ takes the form

$$\Delta U = \frac{N}{\Gamma}\int_0^1 \sqrt{2K_2 l^2 \left(V(u)+\varepsilon_2\right)}\,dt \pm \frac{Nk_1 l^2 \delta_1}{\Gamma} - N\varepsilon_2 + \frac{Nk_1^2 l^2}{2K_1 \Gamma^2}. \qquad (A3)$$

From Eq. (A3), we find the minimum of $\Delta U$ as a function of $\varepsilon_2$. For $\Gamma < N$, the relation between the unit elongation $\delta_1$ and $\varepsilon_2$ corresponding to the minimum energy solution is given by

$$\delta_1 = \mp\frac{1}{k_1 l^2}\int_0^1 \sqrt{2K_2 l^2 \left(V(u)+\varepsilon_2\right)}\,dt \mp \frac{k_1}{K_1 \Gamma}. \qquad (A4)$$

For $\Gamma \geq N$, only one ID can be present in the system. In this case, $\Delta u = \pm 1$ and $\Delta U$ is minimized at $\varepsilon_2 \to 0$.